# A Novel Ensemble Learning Approach for Enhanced IoT Attack Detection: Redefining Security Paradigms in Connected Systems

Hikmat A. M. Abdeljaber[1], Md. Alamgir Hossain[2], Sultan Ahmad[3,4,*], Ahmed Alsanad[5,*] , Md Alimul Haque[6] , Sudan Jha[7] and Jabeen Nazeer[3]

[1] Department of Computer Science, Faculty of Information Technology, Applied Science Private University, Amman, Jordan
[2] Department of Computer Science and Engineering, State University of Bangladesh, South Purbachal, Kanchan, Dhaka 1461, Bangladesh
[3] Department. of Computer Science, College of Computer Engineering and Sciences, Prince Sattam Bin Abdulaziz University, Alkharj 11942, Saudi Arabia
[4] School of Computer Science and Engineering, Lovely Professional University, Phagwara, 144411, Punjab, India
[5] STC's Artificial Intelligence Chair, Department of Information Systems, College of Computer and Information Sciences, King Saud University, Riyadh 11543, Saudi Arabia
[6] Department of Computer Science, Veer Kunwar Singh University, Ara, Bihar, India
[7] Department of Computer Science and Engineering, School of Engineering, Kathmandu University, Banepa, Kathmandu, Nepal

## Abstract

The rapid expansion of Internet of Things (IoT) devices has transformed industry and everyday lives by facilitating widespread connection and data interchange. This increase in connection has generated significant security weaknesses, rendering IoT systems more vulnerable to advanced cyber-attacks. This research introduces a novel ensemble learning architecture focused on at improving the detection of IoT attacks. The proposed approach utilizes advanced machine learning methods, namely the Extra Trees Classifier, and implements intensive preprocessing and hyperparameter optimization to examine datasets including CICIoT2023, IoTID20, BotNeTIoT-L01, ToN_IoT, N-BaIoT, and BoT-IoT. The findings demonstrate remarkable performance, with the model attaining near-optimal metrics, including Recall, Accuracy and Precision, while maintaining very low error rates. These findings demonstrate the model's efficiency above existing techniques, offering an effective choice for securing IoT environments. This research establishes a new benchmark for IoT security, providing a robust basis for future progress in protecting networked devices from emerging cyber threats.

## 1. Introduction

IoT security comprises the techniques and procedures employed to protect IoT devices and networks from cyber threats, illegal intrusions, data breaches, and other unlawful activity. The main objective is to protect the safety, integrity, and accessibility of the data and services these devices supply, ensuring their dependability and trustworthiness [1], [2], [3].

Researching IoT security is essential due to the increasing integration of IoT devices in critical infrastructure, healthcare, industrial automation, and everyday consumer applications [4], [5]. Ensuring the security of these devices is vital to protect sensitive data, maintain user privacy, and prevent disruptions that could have severe economic and societal impacts [6], [7]. By developing advanced detection and mitigation techniques through rigorous research, we can significantly enhance the resilience of IoT systems, foster trust in IoT technologies, and pave the way for their broader and safer adoption in various sectors.

In recent times, there has been a significant rise in IoT-related cyber-attacks, reflecting the rapidly evolving threat landscape. Trends include the increasing frequency of DDoS attacks targeting IoT devices, ransomware campaigns exploiting vulnerabilities in IoT infrastructure, and sophisticated Advanced

Persistent Threats (APTs) aimed at critical IoT-dependent systems. These attacks not only disrupt IoT services but also compromise sensitive data, highlighting the urgent need for robust, adaptable defense mechanisms. Addressing these evolving challenges requires advanced detection models capable of identifying and responding to diverse and complex attack types [8], [9].

Solving the problem of IoT security is particularly challenging in the rapidly advancing tech-based world due to the sheer diversity and scale of IoT devices, which often have limited computational resources and varying security protocols [10]. Additionally, the dynamic nature of cyber threats, with increasingly sophisticated attack vectors and techniques, requires constant vigilance and innovation in defense mechanisms [11], [12]. The heterogeneity of IoT ecosystems, coupled with the need for seamless interoperability and real-time data processing, further complicates the development of robust security solutions. These complexities demand a multifaceted and adaptive approach to effectively safeguard IoT networks against evolving threats [13], [14].

The advantage of the proposed approach over existing methods lies in its novel use of ensemble learning, specifically advancing the Extra Trees Classifier, combined with rigorous data preprocessing and hyperparameter tuning. Traditional approaches, such as signature-based detection and single classifier models, often suffer from high limited adaptability and FPR to new and evolving threats [15]. Moreover, these methods fail to provide consistently accurate results across all types of classifications and datasets. In contrast, our method achieves near-perfect accuracy and reliability across diverse IoT datasets. This ensures not only enhanced detection precision but also significantly reduced error rates. This technique leverages cutting-edge machine learning technologies while emphasizing scalability and flexibility, developing a new standard in IoT security. It gives a better and more complete protection mechanism for dealing with the ever-evolving world of cyber threats. The significant contributions of the research are listed below:

- Evaluating current IoT security methods, highlighting their limitations, such as high false positive rates and poor adaptability, to underscore the improvements offered by our proposed model.
- Introducing an innovative use of the Extra Trees Classifier for IoT attack detection, which significantly enhances the accuracy and reliability of detecting various types of cyber-attacks.
- Evaluating the model rigorously for multiple types of attack classifications across various datasets using multiple evaluation metrics, ensuring comprehensive performance validation.
- This solution is designed to be scalable and adaptable, making it suitable for deployment in various IoT environments, and ensuring robust security against evolving threats.
- The research sets a new benchmark in IoT security, providing a comprehensive defense mechanism that outperforms existing methods.

After the introduction, the paper is organized into several sections. First, it reviews related works. Next, it explains the methodology of the proposed approach. This is followed by the results and discussion. Finally, the paper ends with the conclusion and references.

## 2. Related Work

Researchers have thoroughly explored approaches for identifying IoT security issues. Many deep learning (DL) and machine learning (ML) models have been designed to increase accuracy, efficiency, and dependability. This section evaluates different methods, detailing their benefits and shortcomings. It also gives a thorough comparison to demonstrate advancements in IoT risk identification.

Aldaej et al. [16] used an ensemble technique combining E-Tree, Deep Neural Networks, and Random Forest (RF) to achieve high accuracy across multiple datasets like NSL-KDD, Bot-IoT, CICIDS2018, and IoTID20, though it faced challenges with high computational demands. Similarly, Sham et al. [17] evaluated various machine learning and deep learning models, such as CNN, LSTM, and RF, across datasets including UNSW, CIC-IDS2018, NSL-KDD, NVD, N-BaIoT, and BoT-IoT, showing high detection accuracy but encountering scalability issues.

Yaras and Dener [18] employed a hybrid deep learning approach using LSTM and CNN, achieving high accuracy and robustness in classifying both multiclass and binary attacks on CICIoT2023 and TON_IoT datasets, though their method required significant computational resources. Mohamed et al. [19] focused on ReliefF-based feature selection combined with ML and DL models on the ToN_IoT dataset, yielding efficient feature selection and high accuracy but at the cost of high computational expense. Hairab et al. [20] utilized CNN with regularization techniques to address class imbalance in the TON_IoT dataset, showing robust accuracy but struggled with complex IoT environments.

Khan and Alkhathami [21] employed Random Forest, Adaptive Boosting, and other techniques on the CICIoT2023 dataset, offering efficient feature selection and reduced computational response time, though they encountered difficulties with complex environments. Muñoz and Valiente [22] applied stacked autoencoders with communication graphs on the IoT-23 dataset, providing efficient feature extraction with low computational demands but showing potential for overfitting. Hasan et al. [23] used models such as RF, DNN, MLP, and AdaBoost on CICIoT2023, excelling in binary classification but facing performance issues with complex multiclass setups. Finally, Neto et al. [24] and Wardana et al. [25] explored hybrid ML and ensemble averaging techniques, respectively, demonstrating robust performance across various attack types but facing challenges with computational complexity and potential overfitting.

Recent studies in IoT security detection have employed a variety of ML and deep learning models, each demonstrating significant strengths but also encountering notable limitations. Key challenges include high computational costs, potential overfitting, difficulty in detecting rare attack types, scalability issues, and performance degradation with increasing complexity in multiclass classifications. This proposed ensemble-based Extra Trees classifier effectively addresses and overcomes these limitations, offering a robust and efficient solution for IoT attack detection.

Many recent approaches, such as those utilizing hybrid deep learning models and ensemble techniques, demand substantial computational resources, require significant processing power and memory. The proposed approach mitigates this issue by leveraging the Extra Trees classifier, which is inherently more efficient in both training and prediction phases due to its random selection of features and bootstrap samples. This leads to reduced computational overhead while maintaining high accuracy. The Extra Trees classifier combats overfitting through its ensemble nature, averaging the results of multiple decision trees to generalize better to unseen data. Additionally, the randomization process in tree construction further reduces the likelihood of overfitting, ensuring the model remains robust across diverse datasets.

Detecting rare attack types is a challenge for various existing models. The Extra Trees classifier excels in this aspect by constructing a large number of trees, each considering different subsets of features and samples. This ensemble diversity increases the probability of capturing and accurately classifying rare attack patterns, thereby enhancing the detection capabilities of the model.

Scalability is a significant limitation in many complex ML and DL models. This proposed model scales efficiently with large datasets due to the parallel nature of tree construction in the Extra Trees algorithm. This allows for the effective handling of vast amounts of data, making it suitable for real-world IoT environments where data volume and variety are continually increasing. Most of the models experience performance degradation as the complexity of the classification tasks increases, particularly in multiclass scenarios. The Extra Trees classifier addresses this by maintaining high performance across both binary and multiclass classifications. Its ability to select optimal splits and aggregate predictions from multiple trees ensures consistent and reliable performance, regardless of the classification complexity.

The proposed ensemble-based Extra Trees classifier significantly enhances IoT attack detection by overcoming the primary limitations of recent approaches. It offers a computationally efficient, scalable, and robust solution capable of detecting both common and rare attack types with high accuracy. Additionally, this approach provides lower false positive and better true positive rates for all types of classifications and diverse datasets, ensuring reliable performance across various IoT environments.

## 3. Methodology

In this section, the methodology of the proposed approach is described from dataset loading to the classification approach, including the description of preprocessing and evaluation metrics. Fig. 1 represents the pipeline of the proposed approach, showing each stage of the process.

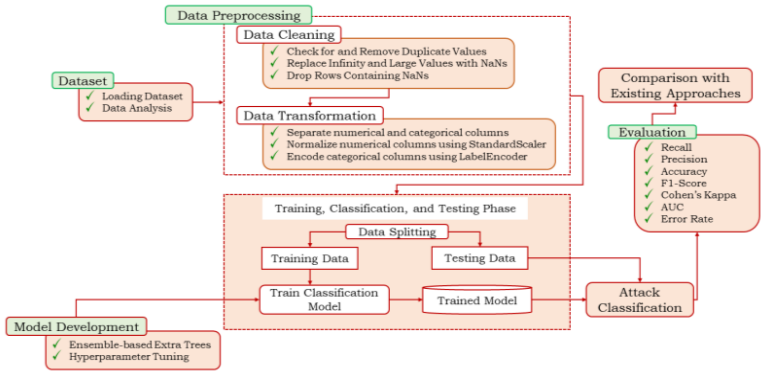

**Fig. 1.** Development Pipeline of the Proposed Approach.

## 3.1 Dataset

The datasets utilized in this research cover a wide range of IoT attack scenarios, providing a comprehensive evaluation environment for the proposed detection model. Table 1 summarizes the different datasets, their respective attack types, and the specific names of the attacks included. The total amount of data before preprocessing is substantial, highlighting the robustness of the evaluation. The CICIoT2023 dataset contains 4,723,822 rows and 47 columns, featuring a variety of attack types such as Mirai, DDoS, DoS, and others. The IoTID20 dataset consists of 625,783 rows and 86 columns, with attacks including Mirai, DoS, Scan, and MITM ARP Spoofing. The BotNeTIoT-L01 dataset includes 2,426,574 rows and 25 columns, distinguishing between Normal and Attack scenarios. The BoT-IoT dataset has 3,668,522 rows and 46 columns, categorizing data into Normal and Bot types. The ToN_IoT dataset has 1,379,274 rows and 14 columns containing injection, ransomware, and others. Lastly, the N-BaIoT dataset comprises 1,018,298 rows and 116 columns, encompassing attack types like Mirai and Gafgyt.

**Table 1.** Used Dataset Scenarios

| Dataset | Attack Type | Name of Attack Types |
|---|---|---|
| CICIoT2023 [16] | 2 | Attack; Benign; |
| | 8 | DDoS; Brute Force; DoS; Mirai; Web-based; BenignTraffic; Spoofing; Recon; |
| | 34 | DDoS-UDP_Flood; DDoS-ACK_Fragmentation; DDoS-ICMP_Flood; Uploading_Attack; Recon-PingSweep; DDoS-TCP_Flood; Backdoor_Malware; DDoS-PSHACK_Flood; XSS; DDoS-SYN_Flood; CommandInjection; DDoS-RSTFINFlood; SqlInjection; DDoS-SynonymousIP_Flood; BrowserHijacking; DoS-UDP_Flood; DictionaryBruteForce; Mirai-greip_flood; DoS-TCP_Flood; DDoS-SlowLoris; DoS-SYN_Flood; BenignTraffic; DDoS-HTTP_Flood; Mirai-greeth_flood; VulnerabilityScan; Mirai-udpplain; DoS-HTTP_Flood; Recon-PortScan; DDoS-ICMP_Fragmentation; Recon-OSScan; Recon-HostDiscovery; DDoS-UDP_Fragmentation; DNS_Spoofing; MITM-ArpSpoofing; |
| IoTID20 [17] | 2 | Anomaly; Normal; |
| | 5 | Mirai; MITM ARP Spoofing; Scan; Normal; DoS; |
| | 9 | Mirai-UDP Flooding; Scan Hostport; Mirai-Hostbruteforceg; MITM ARP Spoofing; DoS-Synflooding; Normal; Mirai-HTTP Flooding; Scan Port OS; Mirai-Ackflooding; |
| BotNeTIoT-L01 [18] | 2 | Normal; Attack; |
| N-BaIoT [19] | 2 | Benign; Bot; |
| | 3 | Benign; Mirai; Gafgyt; |
| | 11 | gafgyt_junk; mirai_udpplain; Benign; mirai_ack; mirai_udp; mirai_syn; mirai_scan; gafgyt_scan; gafgyt_udp; gafgyt_tcp; gafgyt_combo; |
| BoT-IoT [20] | 2 | Normal; Bot; |
| ToN_IoT [21] | 2 | Attack; Benign; |
| | 10 | injection; ddos; Benign; password; xss; scanning; dos; backdoor; mitm; ransomware; |

### 3.2 Data Preprocessing

The preparation of the initial information via preprocessing is a critical step before analysis. This involves cleaning, transforming, and organizing the data. For this research, preprocessing ensures the dataset is appropriately prepared for training the ensemble-based machine learning model. This phase serves as essential for boosting the model's accuracy and efficiency in identifying IoT attacks [22].

At first, the datasets were loaded from the directory. This step involves reading the data files into a structured format suitable for analysis. Each dataset is read as a DataFrame ensuring that the data is accessible for the subsequent preprocessing stages. Next, the data is cleaned to improve its quality and consistency, which is critical for accurate model training. The first sub-step involves removing duplicate values. Duplicate rows are identified and dropped using the *drop_duplicates()* function to ensure that each row in the dataset is unique.

Following this, every instance of infinity and overly large numbers are replaced with NaNs to manage outliers and erroneous data points. Infinity values, represented as ∞, are substituted with NaNs. Lastly, rows with NaN values are deleted to ensure the dataset is filled and acceptable for analysis. These processes together assure the data is clean and suitable for further processing.

After the data cleaning procedures, the data is transformed to ensure it is uniform and acceptable for model training. The initial stage in this procedure is separating numerical and category columns. Subsequently, the numerical columns are normalized using the StandardScaler to guarantee that the data follows a standard normal distribution. The normalization process is defined by:

$$X'_{num} = \frac{X_{num} - \mu}{\sigma} \quad (1)$$

Where $\mu$ is the mean and $\sigma$ is the standard deviation of the numerical data $X_{num}$.

Finally, categorical columns are encoded using the LabelEncoder to convert categorical data into numerical format, facilitating its use in machine learning algorithms. The encoding process is given by:

$$X'_{num} = LabelEncoder().fit\_transform(X_{cat}) \quad (2)$$

Where $X_{cat}$ categorical data.

These transformations ensure that the dataset is standardized and all features are in a numerical format appropriate for training the machine learning model. These procedures are crucial to increasing the model's performance and delivering thorough IoT security analysis.

### 3.3 Classification with Extra Trees

The Extra Trees classifier, also known as Extremely Randomized Trees, is a machine learning method identified for its ability to handle different datasets properly. It works by combining predictions from many decision trees, which increases both accuracy and dependability [23]. Using ensemble learning, this system can handle complicated data patterns and connections. This makes it perfect for tasks such as detecting IoT risks, where reliable categorization and data variability are critical. Its capacity to prevent overfitting and increase generalization makes it a desirable tool for our research. The classification process, as illustrated in Fig. 2, utilizes an ensemble-based Extra Trees classifier to enhance predictive accuracy through majority voting from multiple decision trees, detailed in Algorithm 1.

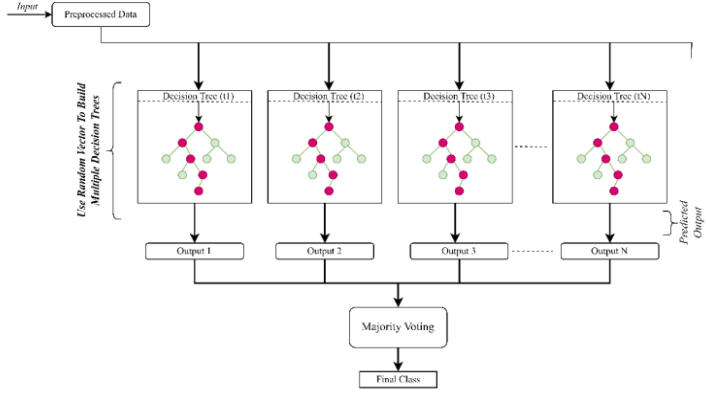

**Fig. 2.** IoT Attack Classification Process with Extra Trees.

**Algorithm 1.** Training Process for Proposed Ensemble-based Extra Trees Classifier

i. Initialize an ensemble of $n$ decision trees $T = \{T_1, T_2, \ldots, T_n\}$.
ii. For each tree $T_i$ in the ensemble:
   Select a bootstrap sample $S_i$ from the training data $(X_{train}, y_{train})$.
   $S_i \{(x_j, y_j) | j \in random\ sample\ of\ \{1, 2, \ldots, N_{train}\}\}$ ; Where, $N_{train}$ is the number of training samples.
iii. Tree Construction:
   For each node in the tree, randomly select $k$ features from the total $m$ features.
   $F_i = RandomSelect(k, m)$; Where, $F_i = \{f_1, f_2, \ldots, f_k\}$.
iv. For each selected feature $f \in F_i$ find the best split point $s$ that minimizes the impurity measure.
   a) Calculate the impurity for a split $s$ on feature $f$:
   $I(s, f) = \frac{N_{left}}{N} I_{left} + \frac{N_{right}}{N} I_{right}$; Where, $N$ is the total number of samples, $N_{left}$ and $N_{right}$ are the number of samples in the left and right child nodes, respectively, and $I_{left}$ and $I_{right}$ are the impurity measures of the child nodes.
   b) The best split $s^*$ for feature $f$ is the one that minimizes $I(s, f)$:
   $s^* = arg\ \min_s I(s, f)$
v. Split the node using the best split $s^*$ and create child nodes. And Repeat steps iii-v recursively for each child node until a stopping criterion is met.
vi. Continue building the tree until all nodes are either pure or meet the stopping criteria.
vii. Aggregate the ensemble of $n$ trained trees $T = \{T_1, T_2, \ldots, T_n\}$ to form the final Extra Trees classifier.

The training process for the Extra Trees classifier involves constructing an ensemble of decision trees using randomly selected features and bootstrap samples from the training data. Each tree is built by recursively selecting the best splits based on impurity measures, ensuring diversity among the trees. The term $s$ in the algorithm represents each possible split point along a given feature $f$, while $s *$ denotes the optimal split point that minimizes the impurity $I(s, f)$. This optimal split selection helps enhance the model's classification performance by reducing impurity at each node, improving its robustness and accuracy.

The final Extra Trees classifier is formed by aggregating the predictions of all trained trees, providing a robust model for various classification tasks.

Firstly, the input data $X_{test}$ is prepared for prediction, ensuring it is preprocessed in the same manner as the training data. For each instance $x_j$ in the test dataset $X_{test}$, the instance is passed through each decision tree $T_i$ in the ensemble $T = \{T_1, T_2, \ldots, T_n\}$. Each tree $T_i$ provides a predicted class label $T_i(x_j)$.

The next step involves aggregating the predictions from all trees for each instance $x_j$. The final predicted class $y_{pred,j}$ is determined by majority voting among the predictions of all $n$ trees:

$$y_{pred,j} = mode(\{T_1(x_j), T_2(x_j), \ldots, T_n(x_j)\}) \quad (3)$$

This majority voting mechanism ensures that the prediction leverages the combined wisdom of multiple trees, thereby enhancing the robustness and accuracy of the classification.

The predicted class $y_{pred,j}$ for each instance $x_j$ indicates the type of attack or benign activity identified by the model. This classification step is crucial for detecting and responding to various IoT system attacks in real-time, providing a robust defense mechanism for the IoT environment.

### 3.4 Evaluation Metrics

This research uses evaluation metrics to thoroughly assess the model's performance in detecting IoT attacks from various perspectives [24]. Table 2 summarizes these metrics.

**Table 2.** Description of Evaluation Metrics

| Metric Name | Description | Equation |
|---|---|---|
| Accuracy | Measures the fraction of accurately predicted instances out of the total instances. | $\frac{TP + TN}{TP + FN + FP + TN}$ |
| Precision | Measures the fraction of actual positive predictions out of the total predicted positives. Indicates the model's ability to prevent false positives. | $\frac{TP}{TP + FP}$ |
| Recall | Measures the fraction of true positive predictions out of the total actual positives. Indicates the model's capacity to recognize all relevant occurrences. | $\frac{TP}{TP + FN}$ |
| F1-Score | The harmonic mean of recall and precision, offering a balance between the two metrics. | $\frac{2 * (recall * precision)}{recall + precsion}$ |
| Cohen's Kappa | Measures the agreement between predicted and actual labels, including for any possibility of agreement resulting by a mistake. | $\frac{P_o - P_e}{1 - P_e}$ |
| AUC Score | The area under the ROC curve, demonstrating the model's ability to differentiate across classes. A greater AUC suggests better performance. | $\int TPR(FPR)dFPR$ |
| Error Rate | Measures the fraction of inaccurate predictions out of the total predictions. | $1 - Accuracy$ |

In Table 3, FN is False Negative, TP is True Positive, FP is False Positive, TN is True Negative, $P_e$ is the expected agreement, and $P_o$ is the observed agreement by chance calculated from the Confusion Matrix.

## 4. Results and Discussion

This section provides the results and discussion of the implemented model, presenting comprehensive evaluations through proper tables and figures. The performance metrics across multiple datasets are analyzed, demonstrating the effectiveness and robustness of the proposed approach in detecting IoT attacks.

The experiments in this research were carried out on a high-performance ASUS device. The system is equipped with an 11th Gen Intel Core i7-11700 processor running at 2.50 GHz and 16 GB of RAM. It operates on a 64-bit Windows 11 Pro system. Anaconda Navigator was used for managing software environments, with Jupyter Notebook as the primary development tool. Key Python libraries such as Pandas, NumPy, Scikit-learn, and Matplotlib supported tasks like data preprocessing and model evaluation. The Extra Trees Classifier was employed, with hyperparameter tuning applied to optimize its performance. Evaluation metrics included recall, accuracy, precision, F1-score, Cohen's Kappa,

AUC, and error rate to thoroughly measure the model's effectiveness in detecting IoT attacks. The system provided the computational power and flexibility needed to address the challenges of IoT security. The datasets were split into 70% for training and 30% for testing, with results based on the test data.

## 4.1 Results of the Model for Various Datasets

Table 3 presents the performance metrics of the proposed model across various datasets, demonstrating its effectiveness in IoT attack detection. The model achieves near-perfect scores for all metrics, and maintains minimal Error Rates, underscoring its robustness and reliability.

**Table 3.** Values of Evaluation Metrics ofor Different Datasets

| Dataset | Classification | Recall | Precision | Accuracy | F1-Score | Cohen's Kappa | AUC | Error Rate |
|---|---|---|---|---|---|---|---|---|
| CICIoT2023 | 2 | 1.00000 | 1.00000 | 1.00000 | 1.00000 | 1.00000 | 1.00000 | 0.00000 |
|  | 8 | 0.99999 | 0.99999 | 0.99999 | 0.99999 | 0.99997 | 1.00000 | 0.00001 |
|  | 34 | 0.99958 | 0.99958 | 0.99958 | 0.99957 | 0.99954 | 1.00000 | 0.00042 |
| IoTID20 | 2 | 1.00000 | 1.00000 | 1.00000 | 1.00000 | 1.00000 | 1.00000 | 0.00000 |
|  | 5 | 0.99996 | 0.99996 | 0.99996 | 0.99996 | 0.99994 | 1.00000 | 0.00004 |
|  | 9 | 0.99345 | 0.99347 | 0.99345 | 0.99346 | 0.99210 | 0.99985 | 0.00655 |
| BotNeTIoT-L01 | 2 | 1.00000 | 1.00000 | 1.00000 | 1.00000 | 1.00000 | 1.00000 | 0.00000 |
| N-BaIoT | 11 | 0.99996 | 0.99996 | 0.99996 | 0.99996 | 0.99995 | 1.00000 | 0.00004 |
|  | 3 | 1.00000 | 1.00000 | 1.00000 | 1.00000 | 1.00000 | 1.00000 | 0.00000 |
|  | 2 | 1.00000 | 1.00000 | 1.00000 | 1.00000 | 1.00000 | 1.00000 | 0.00000 |
| BoT-IoT | 2 | 1.00000 | 1.00000 | 1.00000 | 1.00000 | 1.00000 | 1.00000 | 0.00000 |
| ToN_IoT | 10 | 0.99999 | 0.99999 | 0.99999 | 0.99999 | 0.99999 | 1.00000 | 0.00001 |
|  | 2 | 1.00000 | 1.00000 | 1.00000 | 1.00000 | 1.00000 | 1.00000 | 0.00000 |

These results demonstrate the model's capability to effectively and accurately detect IoT attacks and set a new benchmark in IoT security. The minimal error rates and high Cohen's Kappa values further emphasize the model's reliability and robustness in various IoT environments.

Fig. 3, 4, and 5 illustrate the confusion matrices for the proposed model of CICIoT23 dataset, each representing the classification performance for different attack types. The confusion matrices consistently show high true positive rates and low false positive rates across classifications. These results demonstrate the model's exceptional accuracy and reliability in detecting IoT attacks.

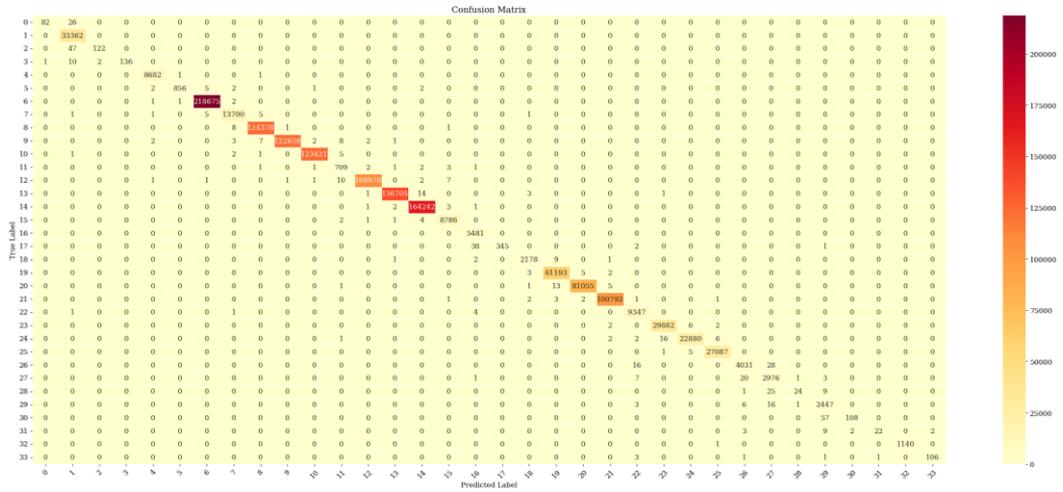

**Fig. 3.** Confusion Matrix for CICIoT2023 Dataset for 34 Classification.

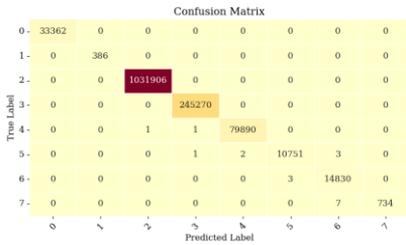

**Fig. 4.** Confusion Matrix for CICIoT2023 Dataset for 8 Classification.

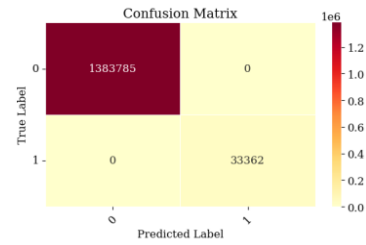

**Fig. 5.** Confusion Matrix for CICIoT2023 Dataset for Binary Classification.

### 4.2 Comparing of the Proposed Model with present Methodologies

Table 4 provides a comparison of the proposed approach with existing methods for the CICIoT2023 dataset across various classifications. The proposed model consistently outperforms other approaches, achieving perfect or near-perfect scores in recall, precision, accuracy, and F1-score across binary, 8-class, and 34-class classifications. This demonstrates the superior performance and reliability of the proposed model in IoT attack detection.

**Table 4.** Comparison of Proposed Approach with Existing Approaches for CICIoT2023 Dataset

| Approach with Reference | Category | Recall (%) | Precision (%) | Accuracy (%) | F1-Score (%) |
| --- | --- | --- | --- | --- | --- |
| Proposed | Binary | 100.0 | 100.0 | 100.0 | 100.0 |
| | 8 Classes | 99.99 | 99.99 | 99.99 | 99.99 |
| | 34 Classes | 99.96 | 99.96 | 99.96 | 99.96 |
| LSTM, 2024 [25] | 34 Classes | 98.75 | 98.66 | 98.75 | 98.59 |
| RF, 2024 [26] | Binary | 99.55 | 99.55 | 99.55 | 99.55 |
| | 8 Classes | 95.54 | 95.55 | 95.54 | 95.51 |
| | 34 Classes | 96.32 | 96.28 | 96.32 | 96.26 |
| MLP, 2024 [27] | Binary | 99.00 | 99.00 | 98.83 | 99.00 |
| | 8 Classes | 97.00 | 97.00 | 97.47 | 97.00 |
| RNN, 2024 [28] | 34 Classes | 96.52 | 96.25 | 96.52 | 95.73 |
| RF, 2024 [29] | Binary | * | 99.45 | 95.64 | 99.45 |
| CFS with BRFC, 2023 [30] | Binary | 95.92 | 95.99 | 99.62 | 95.96 |
| | 8 Classes | 90.29 | 78.25 | 99.43 | 81.86 |
| | 34 Classes | 82.41 | 74.23 | 99.23 | 76.03 |
| BiLSTM, 2023 [31] | 8 Classes | 93.13 | 91.80 | 93.13 | 91.54 |
| DT, 2023 [32] | 34 Classes | 84.14 | 83.71 | 99.23 | 83.81 |
| RF, 2023 [16] | Binary | 96.51 | 96.53 | 99.68 | 96.52 |
| | 8 Classes | 91.00 | 70.54 | 99.43 | 71.92 |
| | 34 Classes | 93.15 | 70.44 | 99.16 | 71.40 |

Table 5 presents a comparison of the proposed approach with other existing methods for the N-BaIoT dataset. The proposed model demonstrates superior performance, significantly outperforming other approaches.

**Table 5.** Comparing of the Proposed Approach with Existing Existing Approaches for N-BaIoT Dataset

| Approach with Reference | Recall (%) | Accuracy (%) | Precision (%) | F1-Score (%) |
|---|---|---|---|---|
| Proposed | 99.99 | 99.99 | 99.99 | 99.99 |
| Ensemble-based DNN, 2024 [33] | 87.31 | 97.21 | 91.41 | 88.48 |
| LDL, 2023 [34] | 86.32 | 98.37 | 83.31 | 84.47 |
| FGOA-kNN, 2023 [35] | 98.73 | 98.07 | 97.04 | 97.87 |
| ER-VEC, 2023 [37] | * | 95.64 | * | * |
| SGDC, 2023 [36] | 98.42 | * | 98.43 | 98.41 |
| BGWO, 2022 [40] | * | 98.97 | * | * |
| CNN-LSTM, 2022 [39] | 89.00 | * | 94.00 | 85.00 |
| WCC and SVM, 2022 [38] | 94.70 | 96.70 | 94.90 | 94.80 |
| LGBA-NN, 2021 [41] | 90.00 | 90.00 | 85.23 | 86.64 |
| RNN, 2021 [42] | * | 89.75 | * | * |

Table 6 presents a comparison of the proposed approach with other existing methods for the BoT-IoT dataset. The perfect performance across all evaluation metrics for the BoT-IoT dataset, significantly outperforming existing methods and establishing itself as a highly effective solution for IoT attack detection.

**Table 6.** Comparing of the Proposed Approach with Existing Existing Approaches for BoT-IoT Dataset

| Approach with Reference | Recall (%) | Accuracy (%) | Precision (%) | F1-Score (%) |
|---|---|---|---|---|
| Proposed | 100.00 | 100.00 | 100.00 | 100.00 |
| DBO-Catboost, 2023 [46] | 96.10 | 96.10 | 96.20 | 96.10 |
| CTGAN with MLP, 2023 [43] | 98.93 | 98.93 | 99.84 | 99.07 |
| SOPA-GA-CNN, 2023 [44] | 97.75 | 98.20 | 97.67 | 97.71 |
| Modified SVM, 2023 [45] | 97.00 | 97.00 | 97.00 | 97.00 |
| SGDC, 2023 [36] | 92.48 | * | 92.28 | 92.37 |
| BTC-SIGBDS, 2023 [47] | * | 94.98 | * | * |
| Fuzzy Interpolation, 2022 [48] | 98.80 | 96.41 | 98.80 | 98.80 |
| FRI, 2021 [49] | 96.00 | 95.40 | 96.00 | 96.00 |
| C4.5, 2020 [50] | 99.99 | 97.62 | 97.63 | 98.79 |

Table 7 provides a comparison of the proposed approach with other existing methods across various datasets, including ToN-IoT, IoTID20, and BotNeTIoT-L01. The proposed approach constantly generates near-perfect or flawless results in recall, precision, accuracy, and F1-score, indicating its excellent performance in IoT threat detection. This demonstrates the resilience and dependability of the proposed model in efficiently identifying IoT threats, exceeding the performance of previous approaches.

**Table 7.** Comparison of the Proposed Approach with Others for Others Dataset

| Dataset | Approach with Reference | Recall (%) | Accuracy (%) | Precision (%) | F1-Score (%) |
|---|---|---|---|---|---|
| ToN-IoT | Proposed | 99.99 | 99.99 | 99.99 | 99.99 |
| | Hybrid (CNN + LSTM), 2024 [51] | 98.75 | 98.75 | 98.75 | 98.75 |
| | Regularized CNN, 2023 [52] | 97.69 | 97.94 | 98.86 | 98.27 |
| | Hybrid CNN and ELM tuned by Sine Cosine Algorithm (SCA), 2023 [53] | 96.65 | 96.65 | 96.60 | 96.62 |
| | 2D-ACNN, 2023 [54] | 90.10 | * | 87.00 | 89.80 |
| | SSW and XGBoost, 2022 [55] | 98.80 | 98.80 | 98.80 | 98.80 |
| | Medium Neural Network, 2022 [56] | 98.20 | 98.39 | 99.60 | 99.70 |
| IoTID20 | Proposed (5 Classes) | 99.99 | 99.99 | 99.99 | 99.99 |
| | Combined ET, DNN, RF [57] | * | 99.00 | 99.04 | * |
| | XGBoost, 2022 [58] | 68.80 | 83.71 | 83.06 | 82.16 |
| | Ensemble ML, 2020 [17] | 87.00 | 87.00 | 87.00 | 87.00 |
| BotNeTIoT-L01 | Proposed | 100.00 | 100.00 | 100.00 | 100.00 |
| | CNN + LSTM, 2024 [59] | 99.80 | 99.86 | 99.87 | 99.83 |

Based on the results and comparisons presented, the proposed model for IoT attack detection demonstrates exceptional performance across a variety of datasets and attack classifications. The model consistently achieves near-perfect or perfect scores, significantly outperforming existing methods[60]. The high Cohen's Kappa values and minimal error rates further emphasize the model's reliability and robustness in diverse IoT environments. These findings confirm that the proposed ensemble learning approach, with its rigorous data preprocessing and hyperparameter tuning, sets a new benchmark in IoT security, offering a highly effective solution for safeguarding interconnected systems against sophisticated cyber threats.

## 5. Conclusion

This research attempts to enhance IoT attack detection using an ensemble-based Extra Trees classifier. It addresses the increasing need for dependable and effective security solutions in IoT systems. The comprehensive data preprocessing processes, the classification technique, and the evaluation metrics applied to evaluate the model's performance are discussed over the course of the research Overcoming major constraints of current models including high computational costs, overfitting, difficulty in identifying rare attacks, scalability problems, and performance degradation in multiclass scenarios, this suggested approach addresses The findings show that the model offers not only great accuracy but also better true positive rates and recall over several datasets and classification approaches. These findings underscore the model's effectiveness and reliability in real-world IoT applications offering a powerful solution for cybersecurity. Conclusively, this research establishes a new benchmark in IoT attack detection, opening the way for more resiliency and adaptable security solutions. Future work will explore integrating real-time data processing and adaptive learning techniques to further enhance the model's capabilities and responsiveness to evolving cyber threats.

Beyond detection, our approach has implications for proactive intrusion prevention and automated threat response. The model's outputs could support real-time threat mitigation, informing adaptive security policies that enhance IoT network resilience against evolving threats. Looking forward, we acknowledge emerging trends in IoT security, such as edge computing and federated learning, which could complement and enhance our approach. Edge computing offers the potential for deploying the model directly at the network edge, thereby improving response times and scalability for real-time detection. Federated learning, meanwhile, can facilitate decentralized model training across IoT devices, preserving data privacy while maintaining robust attack detection capabilities. Future work will explore these adaptations, paving the way for comprehensive, adaptive, and privacy-preserving IoT security solutions.

## Acknowledgements

The authors are grateful to the Deanship of Scientific Research, King Saud University for funding through Vice Deanship of Scientific Research Chairs.

## Author's Contributions

Conceptualization, MAH, HAMA; Investigation and methodology, SA, AA, MA; Project administration, SA, AA; Resources, SJ, JN; Supervision, SA, SJ; Writing of original draft, MAH, HAMA, MA; Writing of the review and editing, SA, AA, SJ; Software, MAH, SJ; Validation, MAH, MA, JN; Funding acquisition, AA, HAMA, SA.

Every author have read and agreed to publish this version of the paper.

## Funding

The authors are grateful to the Deanship of Scientific Research, King Saud University for funding through Vice Deanship of Scientific Research Chairs.

## Competing Interests

The authors confirm that they have no conflicts of interest related to the publication of this paper.